\documentclass[structabstract]{aa}
\usepackage{natbib}
\usepackage{graphicx}

\usepackage{txfonts}

\begin{document}

   \title{A new model for gravitational potential perturbations in disks of spiral galaxies. An application to our Galaxy}

   \subtitle{ }

   \author{T. C. Junqueira,
          J. R. D. L\'epine,
          C. A. S. Braga
          \and 
          D. A. Barros
          }

   \institute{Instituto de Astronomia, Geof\'isica e Ci\^encias Atmosf\'ericas, Universidade de S\~ao Paulo, 
   Cidade Universit\'aria, S\~ao Paulo, SP, Brazil\\
   \email{tjunqueira@astro.iag.usp.br}}

   \date{Received ; accepted }

% \abstract{}{}{}{}{} 
% 5 {} token are mandatory
 
  \abstract 
   {}
  % aim
 {We propose a new, more realistic description of the perturbed gravitational potential of spiral galaxies,
 with spiral arms having Gaussian-shaped groove profiles. The aim is to reach a self-consistent description of the 
 spiral structure, that is, one in which an initial potential perturbation generates, by means of the stellar orbits,
 spiral arms with a profile similar to that of the imposed perturbation. Self-consistency  is a condition for having
 long-lived structures.}
  % methods 
  {Using the new perturbed potential we investigate the stable stellar orbits in galactic disks for galaxies with no bar or with only a weak bar. 
The model is applied to our Galaxy by making use of the axisymmetric
component of the potential computed from the Galactic rotation curve, in addition to other input  parameters 
similar to those of our Galaxy. The influence of the bulge mass on the stellar orbits in the inner
regions of a disk is also investigated.}
  % results 
  {The new description offers the advantage of easy  control of the parameters of the Gaussian profile of its potential. 
 We compute the density contrast between arm and inter-arm regions. We find a range of values for the perturbation amplitude
 from 400 to 800 km$^2 s^{-2}$ kpc$^{-1}$, which implies an approximate maximum ratio of the tangential force to the axisymmetric
force between 3\% and 6\%. Good self-consistency of arm shapes is obtained  between the Inner
Lindblad resonance (ILR) and the 4:1 resonance. Near the 4:1 resonance the response density starts to deviate from
 the imposed logarithmic spiral form. This creates bifurcations that appear as short arms. Therefore the deviation 
 from a perfect logarithmic spiral in galaxies can be understood as a natural effect of the 4:1 resonance. Beyond
the 4:1 resonance we find closed orbits that have  similarities with the arms observed in our Galaxy. In regions
near the center, elongated stellar orbits appear naturally, in the presence of a massive bulge, without imposing
any bar-shaped potential, but only extending the spiral perturbation a little inward of the ILR. This suggests that
a bar is formed with a half-size $\sim 3$ kpc  by a mechanism similar to that of the spiral arms.}
  % conclusions  
 {The potential energy perturbation that we adopted represents an important step in the direction of self-consistency,
 compared to previous sine function descriptions of the potential. In addition, our model produces a  realistic description
 of the spiral structure, which is able to explain several details that were not yet understood.}


\titlerunning{The spiral structure of the Galaxy}

\authorrunning{T. C. Junqueira et al.}

 \maketitle

%________________________________________________________________

\section{Introduction}

Despite its historical merit, the spiral structure theory, proposed by \cite{Lin1964}, has well$-$known limitations. For instance, Lin $\&$ Shu add gas and star 
into a single component, when their dynamics differ in response to the same potential; stars, for example, are not 
affected by pressure gradients. Additionally, real arms are not tightly wound, as assumed in that theory, and the physical properties of
the disk (densities, velocity perturbations) do not have a sine-shaped dependence in azimuth, as argued in the present paper.
Around the same time,  \cite{Toomre1964}, in an opposing point of view, argued that since the largest fraction of matter in the 
disk is in the form of stars, one can, in a first approximation, neglect the gas and work with stellar dynamics. He devoted much of his attention to
understanding why an excess of stars in a given region does not grow without limit, due to the excess of gravitational attraction.
Regardless, his model for the arms focused more on stellar dynamics, considering them as regions with stellar excess of stars, that rotate with 
velocity that does not diverge from the velocity of the disk matter.

\cite{Kalnajs1972} proposed an innovative way of understanding the spiral structure, which was focused solely on stellar dynamics. He
showed that by introducing some degree of organization to the closed stellar orbits, we can generate regions where these orbits 
are crowded. Such an organization is achieved by rotating orbits of increasing size 
by a given angle, relative to previous ones. These regions of high stellar densities look 
like logarithmic spirals and rotate with the required pattern speed to transform the usually open orbits into closed ones. Furthermore, such an
organization of stellar orbits can be achieved via galactic collisions \citep{Gerber1994}.
The great advantage of the Kalnajs model in comparison to Toomre's is that the former's spiral structure is not disrupted by differential rotation; 
in other words, it can be long lived. It is important to note that the search for stable or quasi-stationary solutions 
has traditionally been the aim of dynamical models of disks. This is still a legitimate concern, given that the near-infrared morphology of 
several grand design galaxies reveals the existence of an underlying two-arm structure constituted mainly of old stars 
\citep{Grosbol1998, Block1991, Rix1993, Rix1994, Block1994}. Also, \cite{Lepine2011b} argued that, based on the 
metallicity gradient, the present spiral structure of Galaxy has an age of at least several billion years.

The Kalnajs model is the most promising one, as it gives a correct understanding of the nature of the arms as regions where stellar orbits 
are crowded. However, Kalnajs' model is only a first approximation, since it makes use of stellar 
orbits associated with an axisymmetric potential. In order to reach self-consistency,  one must introduce the effect of the perturbations 
of the potential on the stellar orbits. The perturbations  produce changes in the shapes of the orbits, which in turn change the regions where the orbits are crowded and generate a new perturbation. If at
the end of this cycle we obtain approximately the same density distribution as at the end of the starting one, we say that the model is self-consistent, 
and that the structure will be long lived. Self-consistent models, or at least approximately self-consistent models in terms of arm shapes,
were constructed by \cite{Contopoulos1986}, \cite{Contopoulos1988} and \cite{Patsis1991} for large spiral galaxies, and by \cite{Amaral1996} 
(hereafter A$\&$L) and \cite{Pichardo2003} for the special case of our Galaxy. In the works of Contopoulos and collaborators and of
A\&L, the classical expression for the perturbation $\Phi_{1}$ was used in a frame of reference rotating with constant velocity $\Omega_p$: 

\begin{equation}
\label{potpert} 
{ 
\Phi_{1}(R,\varphi) = \zeta_0\, R\, e^{-\varepsilon_{s} R}\,\cos \left( m\, \frac{ ln\, R }{tan\, i} - 
m\varphi \right),
}
\end{equation}

\noindent with some diversity in the amplitude $\zeta_0$, the behavior of the amplitude with radius (here represented by 
$R\, e^{-\varepsilon_s R}$), the number of arms ($m$), and the pitch angle $i$.
A property of this expression is that if we fix the radius $R$, the variation of the potential as a function of the azimuthal
angle is a sine function. In the work of \cite{Pichardo2003}, the perturbing potential is represented as a superposition
of a large number of oblate spheroids, with their centers distributed along the locii of the spiral arms. This approach does not
give access to a simple analytical expression for the perturbation. 

One  motivation for this work is that it has become evident that the traditional description of  the potential perturbation
in terms of a sine function in the azimuthal direction is not satisfactory. It is simple to estimate the stellar density in a
region where the orbits are crowded and to compare it with the average field, as well as to perceive that the perturbations are much
better described as narrow grooves or channels in the gravitational potential with an approximately Gaussian profile.
We present an analytical description of the perturbing potential, which is realistic when compared to observations and
produces better self-consistency  compared to other models. Strictly speaking, we give priority to the analysis of one
aspect of self-consistency, which corresponds to the shape of the arms or equivalently to the position of the density maxima in
the plane of the galaxy and to their profiles. We employ the expression ``self-consistency of arm shape'' to describe our results,
although several authors that we mentioned above simply refer to ``self-consistency'' for similar analyses.
Other advantages related to the expression proposed in this work for the perturbation are 1) avoiding fluctuations between positive and negative
values for the potential (as produced by a sinusoidal function) over the whole disk for the whole range of parameters, 2) controlling the arm-interarm contrast and, 3) dealing of the thickness of the spiral arms. 
 
The focus in this work is on normal galaxies in which the  structure is dominated by spiral arms,
not by a bar. Naturally, many galaxies present a weak bar in their center, with an extended region of spiral arms beyond the
end of the bar. The present model can be useful to interpret their structure.

In addition to the motivations mentioned above, there has been considerable improvement in recent years in the knowledge 
of the structure of our Galaxy and, in particular, of its rotation curve and pattern speed (discussed, respectively,
in Sects. \ref{rcg} and \ref{mfg}). Therefore it is worth constructing a model that is adequate for comparisons with observations and
other models. We are conscious that a number of authors consider that our Galaxy has a strong bar, so that our model would not
be applicable, but in any case, the comparison with observations may give us clues to this question. Other models that we 
have already mentioned (A$\&$L, Pichardo et al, Contopoulos $\&$ Grosb{\o}l) use the same hypothesis of spiral-dominated structure 
and are directly comparable.

The organization of this paper is as follows. In Sect. \ref{cm}, we review the classical model for the spiral arms. In Sect. \ref{npsa}, we 
present our new model for the spiral arms discuss the required parameters. In Sect. \ref{dcpf}, we deduce the relation between the 
density contrast and the perturbation amplitude. We constrain the perturbation amplitude to a range of values by means of observational evidence for the density contrast. In Sect. \ref{axc} and subsections,  we compute the axisymmetric potential for two different rotation curves. First, we use a 
realistic rotation curve, which is obtained by fitting the data of the Galaxy. The second model uses a ``flat'' rotation curve that is modeled by two 
axisymmetric components (bulge $+$ disk) with different bulge-to-disk ratios in order to better understand the effect of the bulge component on 
the stellar orbits. In Sect. \ref{is}, we present the equations of motion and the integration scheme, and in Sect. \ref{dr}, we derive the density
response. In Sect. \ref{res}, we discuss the adopted corotation radius and the angular velocity of the spiral 
pattern.  The main results are presented in this section, discussed separately for different ranges of 
radii: between the Inner Lindblad Resonance (ILR) and the 4:1 resonance, beyond the 4:1 resonance, and the connection between the bulge and the 
inner stellar orbits. In Sect. \ref{con}, we give our conclusions.    

%__________________________________________________________________

\section{The spiral arms}

\subsection{The classical model}
\label{cm}

The surface density of a zero-thickness disk can be represented mathematically as the sum 
of an axisymmetric or unperturbed surface density $\Sigma_{0}(R)$, and a perturbed surface density
$\Sigma_{1}(R,\varphi)$, which represents the spiral pattern in a frame that rotates at angular speed 
$\Omega_p$. The azimuthal coordinate at the rotating frame of reference is $\varphi = \theta-\Omega_{p}t$, where
$\theta$ is the angle at the inertial frame. As usual, the physical surface density is given by the real part of the following equation 
\citep[see][]{Binney2008}:

\begin{equation}
\label{dens_1}
{ 
\Sigma_{1}(R,\theta-\Omega_{p}t) = \Sigma_se^{i[m(\theta-\Omega_{p}t) + f_{m}(R)]},
}
\end{equation}

\noindent where $f_{m}(R)$ is the shape function, which describes the spiral, and $\Sigma_s$ is a varying function of radius that gives the 
amplitude of the spiral pattern. From $\Sigma_1$, as described by Eq. \ref{dens_1}, we derive the potential $\Phi_d$, using 
Gauss' law, and obtain

\begin{equation}
\label{pot1}
{ 
\Phi_{1}(R,\theta-\Omega_{p}t) = \Phi_de^{i[m(\theta-\Omega_{p}t) + f_{m}(R)]},
}
\end{equation}

\noindent where

\begin{equation}
\label{phi_sigma}
{ 
\Phi_d = -\frac{2\pi G\Sigma_{s}}{|k|}.
}
\end{equation}

In the above expression, $k$ is the wavenumber and $G$ the Gravitational constant. 
Using Eqs. \ref{pot1} and \ref{phi_sigma}, we get

\begin{equation}
\label{pot2}
{ 
\Phi_{1}(R,\theta-\Omega_{p}t) = -\frac{2\pi G\Sigma_{s}}{|k|}
e^{i[m(\theta-\Omega_{p}t) + f_{m}(R)]}.
}
\end{equation}

\noindent This is a well$-$known result, found by \cite{Lin1969} and derived from the WKB theory (see also Appendix B).

\subsection{The new potential for the spiral arms}
\label{npsa}

Before we begin, it is important to clarify what we understand by spiral arms. The model we adopt is based on the 
idea of  \cite{Kalnajs1972} of stellar orbits of successively increasing radii in the disk, organized in such a way
that they get close together in some regions, thus presenting excesses of stellar density.
 
Traditionally, the perturbed surface density has been described by Eq. \ref{dens_1}. This approach, however, is
not realistic. The brightness profiles observed in galactic disks in circles around the center are not sine functions,
nor are the density profiles obtained theoretically from the crowding of stellar orbits. The surface density we
consider to be realistic is given by a density excess that follows a logarithmic spiral, with a Gaussian profile
in the transversal direction. The potential that produces this kind of density has the form

 \begin{equation}
\label{pp2}
{ 
\Phi_{1}(R,\varphi,z) = -\zeta_0 Re^{-\frac{R^2}{\sigma^2}[1 - \cos(m\varphi -  f_m(R))] 
- \varepsilon_s R -|kz|},
}
\end{equation} 

\noindent  where $\zeta_0$ is the perturbation amplitude, $\varepsilon_s^{-1}$ the scale length of the spiral,
 $\sigma$ the width of the Gaussian profile in the galactocentric azimuthal direction, $ k= \frac{m}{R\tan(i)}$ the wavenumber, and  $f_{m}(R)$ the shape function. The scale-length of the disk and of the spiral are 
 not necessarily the same. This question is discussed in Sect. \ref{dcpf}. The shape function is

\begin{equation}
\label{sf}
{ 
f_m(R) = \frac{m}{\tan(i)}ln(R/R_i) + \gamma,
}
\end{equation}

\noindent where $m$ is the number of arms, $i$ is the pitch angle,  $R_i$ is the point where the spiral crosses
the coordinate $x=0$, and $\gamma$ is only a phase angle.

Solving Poisson's equation using Eq. \ref{pp2}, with the assumption of zero-thickness disk and
of tightly wound spiral arms (TWA) in the plane $z=0$ we have

\begin{equation}
\label{dens}
 {
\Sigma_s = \Sigma_{s_0}e^{-\frac{R^2}{\sigma^2}[1 - \cos(m\varphi -  f_m(R))]},
}
\end{equation} 

\noindent where $\Sigma_{s_0}$ is associated with the perturbation amplitude (see Appendix \ref{apB} for more details) by the equation

\begin{equation}
\label{maxsig}
{
\Sigma_{s_0}(R) = \frac{\zeta_0 m}{2\pi G}\frac{R^2}{\sigma^2|\tan i|}e^{-\varepsilon_s R}.
}
\end{equation}

\begin{figure}
\hspace{-1cm}
\centering
\includegraphics[scale=1.]{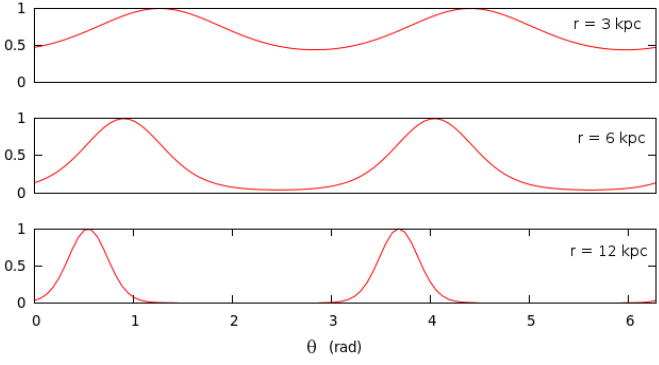}\centering    
\caption{Azimuthal density profile at different radii for $m = 2$, $i = 14^o$, and $\sigma = 4.7$ kpc. Here we set $\Sigma_{s_0}$ = 1, 
since we only want  to show how the density profile varies with the azimuthal angle.}
\label{fig1} 
\end{figure}

\begin{figure}[!ht]
\flushleft 
\parbox{\linewidth}{ 
\begin{center} 
\hskip -0.2cm 
\includegraphics[width=0.5\linewidth]{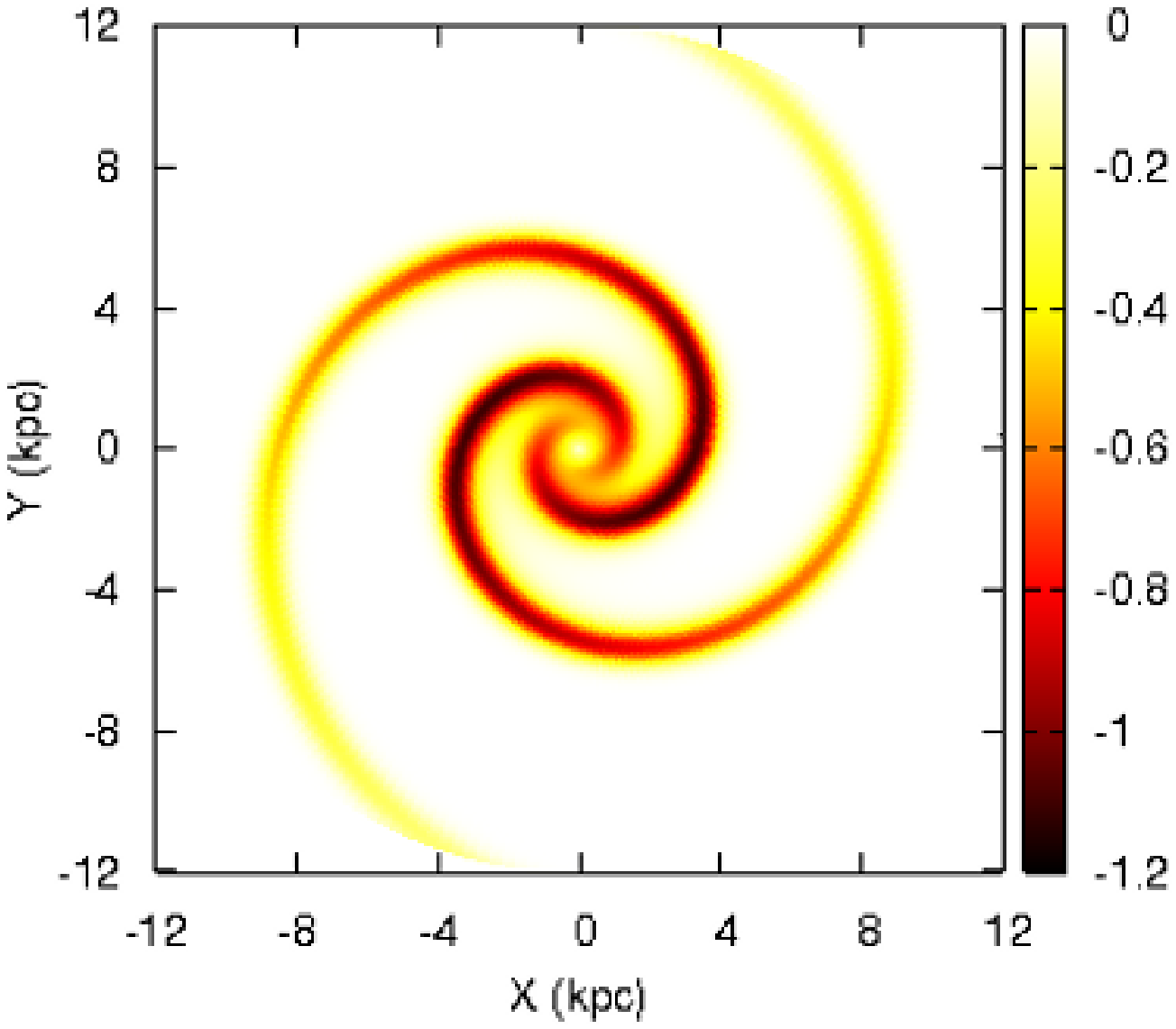}
\hskip 0.0cm  
\includegraphics[width=0.5\linewidth]{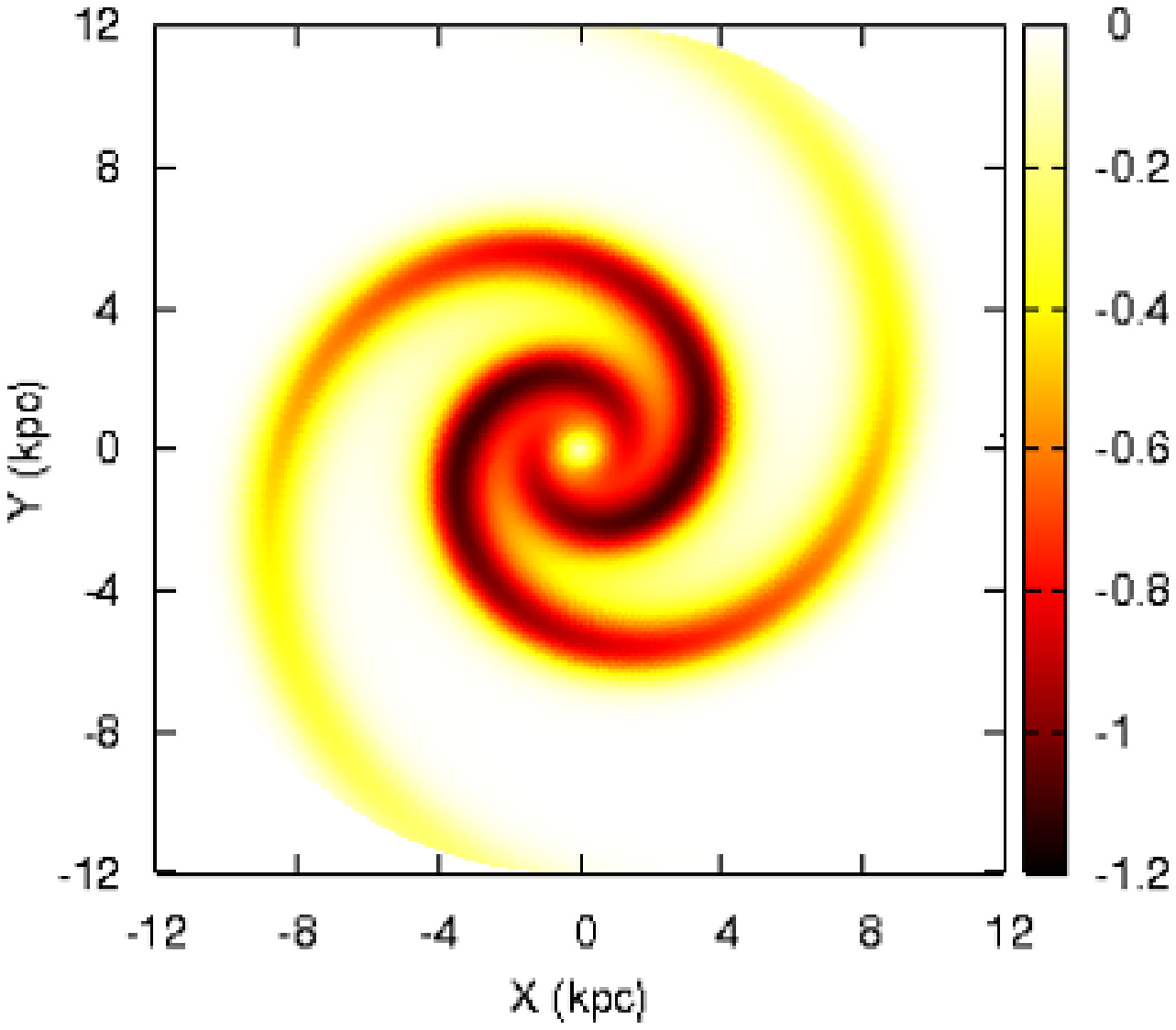}
\end{center}}
\caption{Map of the perturbing potential in the plane of the galaxy.
The colors represent the values of $\Phi_1(R,\varphi)$ in arbitrary units. The picture on the left is the potential 
with $\sigma = 2.5$ kpc and on the  right $\sigma = 4.7$ kpc. For $i = 14^o$ we have on the left 
$\sigma_\perp = 0.6$ kpc and on the right $\sigma_\perp = 1.1$ kpc}
\label{fig2} 
\end{figure} 

\begin{figure}
\hspace{-1cm}
\centering
\includegraphics[scale=0.7]{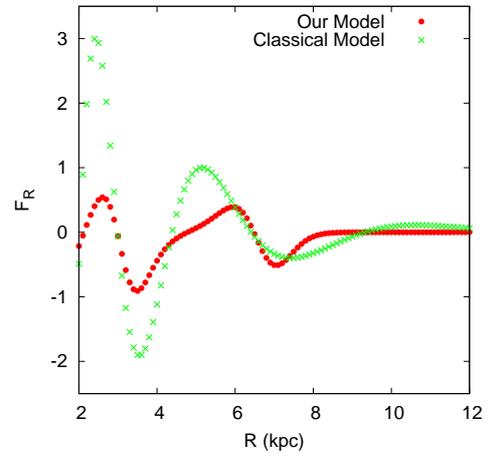}\centering    
\caption{Radial force due to the spiral arm in the direction $\varphi = 0$. The circles in red represent the radial force due to our potential,
while the crosses in green represent the radial force associated with the classical sine function perturbation.  Here, positive values of the force mean that a star would be 
pushed outward and negatives values that it would be pulled inward with respect to the galactic center. In this figure, the number of 
arms is $m=2$, the pitch angle $i=14^o$, $\varepsilon_s = 0.4$ kpc$^{-1}$, and for our model the arm width $\sigma = 4.7$ kpc}
\label{fig3} 
\end{figure}

The azimuthal profile corresponding to Eq. \ref{dens} is illustrated in Fig. \ref{fig1}.
We emphasize that $\sigma$ is the half-width of the spiral arms in directions of galactocentric circles
(see deduction in Appendix \ref{dop}). Because all these  circles  cross the logarithmic arms with
the same angle $i$, the true  width in a direction perpendicular to the arms is given by $\sigma_{\perp}$ 

 \begin{equation}
\label{taw}
{ 
\sigma_\perp = \sigma \sin (i).
}
\end{equation}

\noindent This parameter allows us to reproduce arms of different widths, as shown in Fig. \ref{fig2}. 

Figure \ref{fig3} plots the radial force produced by the spiral arms in our model \citep[red circles, which are very similar to][]{Pichardo2003}  
compared with the classical model (green crosses, derived from Eq. \ref{pot1}) as a function of Galactocentric distance R. In both cases the perturbation amplitude was set to be equal to one, since we just want to see how the force 
profile varies and are not interested in absolute values. The first thing we can notice in Fig. \ref{fig3} is that for the 
classical model the radial force is stronger for the same perturbation amplitude, mainly in inner regions. Thus, the perturbation amplitude 
in the classical case would be lower than in our model. This is due to the fact that a sinusoidal potential varies between 
positive and negative values and that the force, proportional to the variations of potential, is stronger than in a case where the potential
ground level is equal to zero. Another difference between the two models occurs around $9$ kpc, where, in the classical model, the
force becomes positive while in our model the force slowly tends to zero, but it is still negative. This happens because a star in that region
feels the effect of  the last piece of the arm, as we can see in Fig. \ref{fig2} in the $x$ direction, so that it would be pulled inwards.

\section{Relation between density contrast and perturbation amplitude}
\label{dcpf}

Following \cite{Antoja2011} we take as a measure of the density contrast

\begin{equation}
\label{cd}
{ 
A_2 \sim \frac{\Sigma_{s_0}}{\Sigma_d},
}
\end{equation}

\noindent where $\Sigma_d$ is the axisymmetric surface density and $\Sigma_{s_0}$ is the maximum density of the spiral arms. This 
relation is valid under the assumption of a mass-to-light ratio of the order of 1 when $A_2$ is measured in the
infrared bands \citep{Kent1992}. \cite{Antoja2011} collected in the literature  the estimations of 
Galactic and extragalactic  density contrasts, and found them to be in the range $0.13 \leq A_2 \leq 0.23$.
Therefore, an average value would be about  $A_2 = 0.18$.

\begin{table}[!ht]
 \caption{Adopted spiral arms properties}\label{tab:densitycontrast}
\label{tab1}
\begin{tabular}{lccc}\hline
Property & Symbol & Value & Unit \\
\hline
Number of arms & m  & 2 & - \\
Pitch angle & $i$  & $ 14^{o}$ & -  \\
Half width & $\sigma$ & 4.7 & kpc \\
Scale length & $\varepsilon_{s}^{-1}$ & 2.5 & kpc \\
Spiral pattern speed & $\Omega_p$  & 23 & km s$^{-1}$ kpc$^{-1}$   \\
Perturbation amplitude & $\zeta_0$ & 600 & km$^2$ s$^{-2}$ kpc$^{-1}$ \\
\hline
\end{tabular}
\end{table}

The maximum density of the spiral arms is given by Eq. \ref{maxsig}. We refer to the maximum 
at a certain radius, not a maximum over the whole disk. To simplify the analysis, we assume that the axisymmetric surface density
has an exponential behavior (Eq. \ref{sig3}). Galactic disks are often represented by exponential density laws and, depending on 
the value of $\varepsilon_d$, can explain  relatively flat rotation curves. We know that such an exponential  law will
not generate an exact fit to the rotation curve, given by Eq. \ref{galcurve}, but it simplifies the analysis and gives us a
good hint about the value of the perturbation amplitude  

\begin{equation}
\label{sig3}
{ 
\Sigma_d (R) = \sigma_0 e^{-\varepsilon_d R}.
}
\end{equation}

\noindent Therefore the density contrast is

\begin{equation}
\label{dc}
{ 
 \frac{\Sigma_{s_0}}{\Sigma_d} = \frac{\zeta_0 m}{2\pi G}\frac{R^2}{\sigma^2|\tan i|\sigma_0}e^
{-(\varepsilon_s - \varepsilon_d) R}.
}
\end{equation}

The density contrast depends not only on the perturbation amplitude $\zeta_0$, but also 
 on $\sigma$, which describes the arms width. In our model, we can set $\sigma$ as a constant or a 
function of radius.
If we assume that the scale length is the same for both components ($\varepsilon_d = \varepsilon_s$, which is 
a reasonable assumption because we do not want the amplitude of the spiral to drop too fast outwards) and 
compute the value of $\sigma_0$ using the density of the solar neighborhood, $\Sigma_d(R_0) = 49$ M$_{\odot}$ pc$^{-2}$ 
\citep[see][Table 1.1]{Binney2008}, for $R_0 = 7.5$ kpc and $\varepsilon_d^{-1} = 2.5$ kpc we get $\sigma_0 = 984$ $M_{\odot} pc^{-2}$. 
Then, using the values from  Table \ref{tab1} for the number of arms and the pitch angle, Eq. \ref{dc} reduces to

\begin{equation}
\label{contrast2}
{ 
\frac{\Sigma_{s_0}}{\Sigma_d} = 3.10^{-4} \frac{\zeta_0 R^2}{\sigma^2}.
}
\end{equation}

Since we have used, up to now, parameters of the spiral structure that are based on studies of our Galaxy,
we shall estimate the contrast for a galactocentric distance of about 5 kpc, which in our model is about
midway between the ILR and corotation, as discussed in Sect. \ref{res}. Numerically we have 
 $R \approx \sigma$, so that Eq. \ref{contrast2} becomes

\begin{equation}
\label{contrast}
{ 
\frac{\Sigma_{s_0}}{\Sigma_d} \approx 3.10^{-4} \zeta_0.
}
\end{equation}

\begin{figure}[!ht]
\begin{center}
\includegraphics[scale=0.7,angle=0]{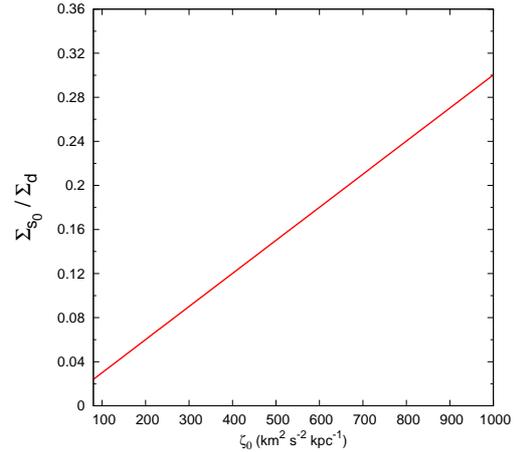}
\caption{Density contrast $\Sigma_{s_0}/\Sigma_d$ as a function of perturbation amplitude for the Milk Way.} 
\label{fig4}
\end{center}
\end{figure}

A plot of the density contrast based on this equation is shown in Fig. \ref{fig4}. Comparing the values
of $A_2$ from the literature, as discussed above, with this figure, we find a range of amplitudes of the perturbation between 400 to 
800 km$^2$ s$^{-2}$ kpc$^{-1}$, which is compatible with the range of density contrasts of Table 1. Taking an average value 
$A_2 = 0.18$  gives us $\zeta_0 = 600$ km$^2$ s$^{-2}$ kpc$^{-1}$.

This is an estimate for $R \approx 5 kpc$. However as we can see from Eq. \ref{contrast2}, the density contrast 
increases with radius. \cite{Kendall2011} and \cite{Elmegreen2011} showed many galaxies with a growth of density contrast
with radius. If we adopt $\varepsilon_d = \varepsilon_s$, the density contrast varies as the square of the radius,
which seems  a little too fast in many cases. This problem can be attenuated in two ways: 1) the scale length of the spiral and of
the disk may be slightly different ($\varepsilon_d < \varepsilon_s$), which will produce an exponential decrease,
2) the arms width may increase with radius, producing
a density contrast with moderate increase, closer to the results of Kendall et al.
For a specific galaxy, it would be possible to check the azimuthal density profile and its variation with radius
and to adjust the parameters to match the observations. One merit of our model is that it facilitates
such adjustments.

\section{The axisymmetric component}
\label{axc}

The main dynamical information of a spiral galaxy is derived from its observed rotation curve. In fact, the rotation
curve gives us the radial gradient of the potential, since we can compute $\Phi_0(R)$ by integrating 
Eq. \ref{force}:

\begin{equation}
\label{force}
{ 
F = -\frac{\partial \Phi_0}{\partial R} = -\frac{V_{rot}^2(R)}{R},
}
\end{equation}

\noindent where $V_{rot}$ is the rotation curve. 

\subsection{The rotation curve of the Galaxy}
\label{rcg}

To contribute to a deep understanding of the spiral structure of any given galaxy, we must make the effort
of introducing the real parameters, such as the dimension of the components, the rotation curve, and the pattern
rotation speed. Although there may be controversies concerning some of these parameters, we chose to illustrate our model of spiral perturbation by applying it to our Galaxy.
However, we also used a simplified analytical expression for the rotation curve, introduced in the
next subsection, with the aim of reaching qualitatively a general understanding of the effect of the the mass of the bulge
on the inner orbits.

\begin{figure}[!ht]
\begin{center}
\includegraphics[scale=0.6,angle=0]{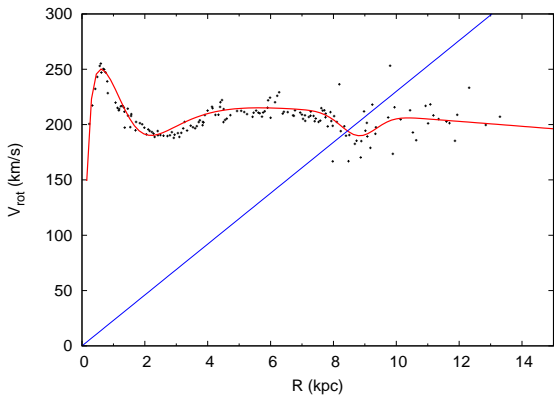}
\caption{Rotation curve of the Galaxy. The red line is the rotation curve fitted using Clemens' data (black dots),
and the blue line is the pattern speed $\Omega_p \times R$  corresponding to an  angular velocity  $\Omega_p = 23$ km s$^{-1}$ kpc$^{-1}$. 
The corotation radius is at 8.4 kpc. We adopted $R_0 = 7.5$ kpc and $V_0 = 210$ km s$^{-1}$.} 
\label{fig5}
\end{center}
\end{figure}

The ``real'' rotation curve of the Galaxy that we have adopted (Fig. \ref{fig5}) is quite flat, except for a local minimum at about $8.8$ kpc.
The existence of this dip in the curve is revealed, for instance, by the study of the epicycle frequency in the galactic disk
\citep{Lepine2008}. This is confirmed in a recent work by \cite{Sofue2009}, which makes use of different data sets, including
data on masers with precise distances measured by parallax, to trace the curve. The rotation curves presented in these two
papers are quite similar, except for minor scaling factors due to the adopted galactic parameters, the solar distance from the Galactic centers
$R_0$, and the solar velocity $V_0$.
The curve is conveniently fitted by two exponentials and a Gaussian

\begin{eqnarray}
\label{galcurve}
V_{rot}(R) = 398e^{-R/2.6 - 0.14/R} + 257e^{-R/65 - (3.2/R)^2}\nonumber 
\\ - 20e^{-[(R - 8.8)/0.8]^2}. 
\end{eqnarray}

\begin{figure}[!ht]
\begin{center}
\includegraphics[scale=0.8,angle=0]{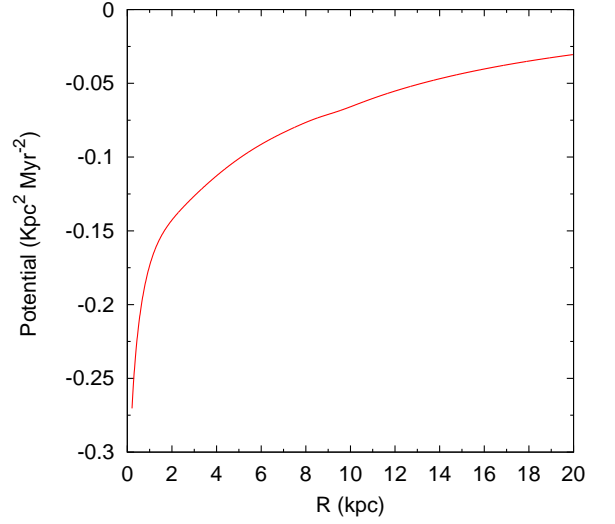}
\caption{Axisymmetrical potential as a function of the galactocentric radius R.} 
\label{fig6}
\end{center}
\end{figure}

The first two components contain terms in $(1/R)$ and $(1/R)^2$ inside the exponential function, 
which produce a decrease of their contribution towards small radii. The last (Gaussian) term represents the local dip.
The interpretation of a similar curve, except for the dip, in terms of components 
of the Galaxy is given by \cite{Lepine2000} and is different from that of \cite{Sofue2009}. The pattern speed is
discussed in Sect. \ref{mfg}. 

Figure \ref{fig6} shows the axisymmetric potential that results from the integration
of Eq. \ref{force}. We used the trapezium rule with adaptive step to solve the integral, and we 
fixed the potential as equal to $\Phi_0(100) = 0$ at R $=$ 100 kpc in order to set the arbitrary constant.

\subsection{A model of a flat rotation curve with a bulge}

Many galaxies have a flat rotation curve, but some of them present a peak 
in the inner part. For instance, as seen in Fig. \ref{fig5}, a peak for our Galaxy appears at a radius
smaller than 2 kpc, with a maximum around 300 $pc$. The nature of the peak is a subject of debate. Some 
authors consider  that it is not due to rotation, like \cite{Burton1993}, who believe that there is
a strong velocity component of gas outflow  from the central regions of the Galaxy. Others 
consider that an important departure from axisymmetric, like a triaxial bulge according to \cite{Gerhard1986},
is required to explain the peak. To perform a first approximation analysis on the effects of the presence of a bulge on the stellar orbits in the inner regions of a galaxy,
we adopt the hypothesis that the bulge is axisymmetric, as it seems to be, at least approximately, in most
spiral galaxies  \citep{Mendez-Abreu2010}. Also in axisymmetric bulges, one can find elliptical orbits
in their central regions. Then, even a very small perturbation could give rise to an oval structure.
Since in this part of the work we are not interested in solving the specific case of our Galaxy, but want to have
a more general idea of the effect of a bulge, we do not use the observed rotation curve of our Galaxy. Instead, we use a simpler
one from which the potential is obtained analytically, thus avoiding numerical integrations. For this study
we adopt the model for the rotation curve given by \cite{Contopoulos1986},  in which the presence of the peak is
modeled by two axisymmetric components (bulge + disk):

\begin{equation}
\label{flatcurve}
{ 
V_{rot}(R) = V_{max}\sqrt{f_{b}\epsilon_{b}R\exp(-\epsilon_{b}R) + [1-\exp(-\epsilon_{d}R)]}.
}
\end{equation}

Here $\epsilon_{b}^{-1}$ and $\epsilon_{d}^{-1}$ are the scale length for the bulge and disk components,
respectively. The relative importance of the two components is given by the bulge fraction $f_b$.
   
We analyzed the inner stellar orbits obtained with rotation curves with different values of $f_b$, that is, with 
different importance of the bulge.  The results are  discussed later, in Sect. \ref{res}.   
   
\section{Stellar orbits}
\label{so}

\subsection{The integration scheme}
\label{is}

\begin{figure}[!ht]
\begin{center}
\includegraphics[scale=0.2,angle=0]{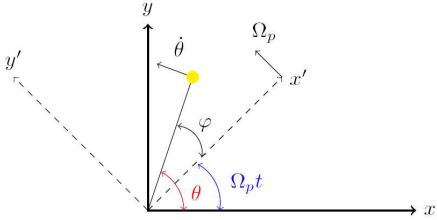}
\caption{Scheme of the Galactic plane. The yellow dot represents a star rotating with speed $\dot{\theta}$ in a
frame of reference (dashed line) corotating with  the spiral arms at the angular velocity $\Omega_p$.} 
\label{fig7}
\end{center}
\end{figure}

Once we know the two components of the potential, $\Phi_0(R)$ and $\Phi_{1}(R,\varphi)$, we can derive the
equations of motion governed by them. Figure \ref{fig7} illustrates the angles and angular velocities used in the
equations. When the Galactic plane is represented, the motion of stars and the pattern speed are shown 
clockwise, which corresponds to negative values of angular velocities. 

The Hamiltonian for such a system is given by

 \begin{equation}
\label{HJ}
{
H=\frac{1}{2}\left[{p_r}^2 + \frac{{J_1}^2}{R^2}\right]-\Omega_{p}J_1 + \Phi_0(R) + \Phi_{1}(R,\varphi), 
}
\end{equation}

\noindent where p$_r$ and $J_1$ are the linear and angular momentum per mass unit, respectively. It is important to note that
$J_{1}$ is measured with respect to the inertial frame. The equations of motion are 

\begin{equation}
\label{drdt}
\frac{dR}{dt}=p_r, 
\end{equation}

\begin{equation}
\label{dpdt}
\frac{dp_r}{dt}=\frac{J_{1}^2}{R^3}-\frac{\partial{\Phi_{0}(R)}}{\partial{R}}
-\frac{\partial{\Phi_{1}(R,\varphi)}}{\partial{R}},
\end{equation}

\begin{equation}
\label{dfdt}
\frac{d\varphi}{dt}=\frac{J_1}{R^2}-\Omega_p,
\end{equation}

\begin{equation}
\label{dJdt}
\frac{dJ_1}{dt}=-\frac{\partial{\Phi_{1}(R,\varphi)}}{\partial{\varphi}}.
\end{equation}

The equations above were integrated using the $6^{th}$ order implicit Runge-Kutta-Gauss (RKG) \citep[see][for
more details]{Sanz1994}, with a fixed time step $dt = 10^{-1}$ $Myrs$. To find the stable periodic orbits, we applied the method of 
Poincar\'e's surface of section \citep{Poincare1892}. The Poincar\'e section was fixed where the orbits cross the axis $\varphi$ = 0$^o$. 
The periodic orbits trapped around circular orbits are represented by a sequence of points in the phase-space (R,p$_r$), lying on a curve 
called an invariant curve.

In practice, what we do is to compute the energy of a circular orbit using Eq. \ref{HJ}. Thus this equation becomes

\begin{equation}
\label{en}
{
h(R)=\frac{{J_0}^2}{2R^2} -\Omega_{p}J_0 + \Phi_0(R). 
}
\end{equation}

This expression has no perturbation terms. The angular momentum per mass unit is given by $J_0 = R V_{rot}$, and $\Phi_0(R)$ is the galactic 
potential. For each radius, we have a value of energy $h$, then we come back to the Eq. \ref{HJ} with $H = h$. Thus, we have to find
the initial conditions $(R,p_{r},J_1,\varphi)$ that will close the orbit for a given $h$. The last variable can be fixed at $\varphi = 0$, 
and the angular momentum $J_1$ is found solving Eq. \ref{HJ} for a given $R$ and $p_r$. Therefore,  we only have to deal with the pair 
of variables $(R,p_r)$. The pair that gives us a stable periodic orbit is found at the center of the island in the Poincar\'e section.

The families of periodic orbits that we obtained (starting from energies equivalent to circular orbits spaced $0.2$ kpc
apart, from 3 up to 8 kpc) are shown in Fig. \ref{fig8}. The inner part from 2 up to 3 kpc was avoided here;
we discuss this region in Sect. \ref{bb}.

\subsection{Response density}
\label{dr}

The response density is computed by considering the conservation of flux between two successive orbits in the 
perturbed and unperturbed cases \citep{Contopoulos1986}. In our case we consider a series of circular orbits spaced at $0.2$ kpc. We can imagine in each orbit a large number of stars equally spaced in the unperturbed case.   
The area $S$ of an angular sector, which is between two neighboring orbits and two successive stars in each orbit,  is transformed
into $S'$ when the perturbation is introduced. The two areas contain the same mass, so that the density 
is proportional to $1/S'$. There are two reasons for the change in distance between 
neighboring stars. In the radial direction, it is because the orbits approach or recede one from the other.
In the azimuthal direction, it is because the stars travel at different angular velocities along the orbit.
Due to the conservation of angular momentum, the stars move more slowly when they are at larger galactic radius, and get closer to one another, since the flux is conserved. This is easy to take into account, since 
the spacing between stars is     $\Delta \varphi \propto \dot{\varphi}^{-1}$.
The response density can be written as

\begin{equation}
\label{densr}
\Sigma_{resp} =\frac{ \Sigma_c 2 \pi R_c \Delta R_c\Delta t}{T R \Delta R
\Delta \varphi},
\end{equation}

\noindent where $\Sigma_c$ is the density, $R_c$ is the radius, and $\Delta R_c$ is the spacing
between two successive orbits, all of them in the unperturbed case, while the same quantities 
in the denominator refer to the perturbed situation.
%%%%%%%%%%%%%%%%%%%%%%%%%%
The quantity $\Delta t$ is the time spent by 
a star to move from $\varphi$ to $\theta$ + $\Delta \varphi$, while $T$ is the period.
In practice, what we do is to divide the circles in N sectors, thus making it possible to find the positions
$R$ and the time for each angle, defined by $i\Delta \theta$ $(i = 1..N)$. Therefore we can find the time 
$\Delta t$ spent by the star to move between two adjacent sectors.

Since we are interested in the position of maximum response density for each orbit, Eq.
\ref{densr} reduces to

\begin{equation}
\label{densr2}
\Sigma_{resp} \propto \frac{ \Delta t}{R\, \Delta R\, \Delta \theta}.
\end{equation}

As the values of $\Sigma_c$, $R_c$, $\Delta R_c$, and $T$ do not change in a given circle, they can
be ignored in Eq. \ref{densr}. Thus the position of maximum density   is the sector where
Eq. \ref{densr2} reaches its maximum value.
%%%%%%%%%%%%%%%%%%%%%%%%%%%%%%

\begin{figure}[!ht]
\begin{center}
\includegraphics[scale=0.7,angle=0]{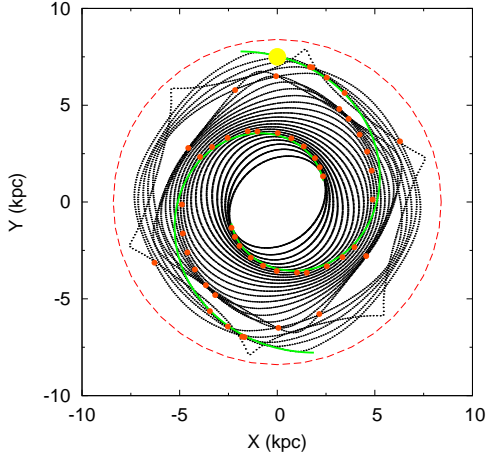}
\caption{Here we show a series of closed orbits in the plane of the Galaxy obtained using the rotation curve of Eq. \ref{galcurve}. 
The spirals indicated in green represent the imposed perturbation; they coincide with the response density (red dots) up 
to 4:1 resonance. The red circle is the corotation radius ($R_{cor} = 8.4$ kpc). The perturbation amplitude 
is $\zeta_0 = 600$ km$^2$ s$^{-2}$ kpc$^{-1}$.
The yellow dot shows the Sun position.} 
\label{fig8}
\end{center}
\end{figure}

We are interested in the position of maximum response density for each orbit. 
The position of the density maxima is shown in Fig. \ref{fig8}. 

\begin{figure}
\hspace{-1cm}
\centering
\includegraphics[scale=0.65]{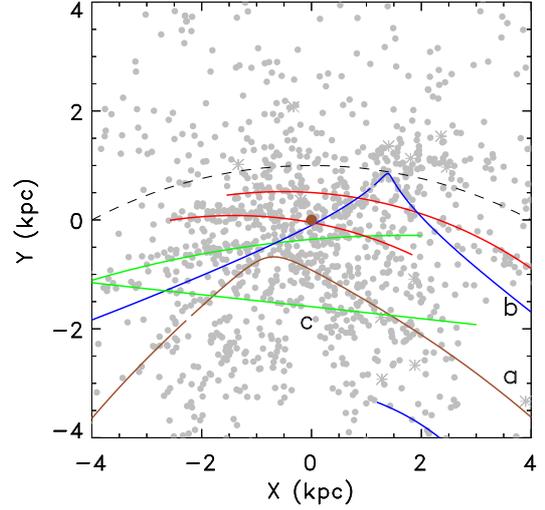}\centering    
\caption{This figure shows a plot of the observed arm structure of the Galaxy in the solar neighbourhood. The Sun is at the center of the
figure, and the distances on the two axes are relative to the Sun. The Galactic center is downward and outside 
the figure. Different kinds of objects have been added to better trace the structure, all of them  shown with 
a same symbol (gray dots): Cepheids, open clusters, CS sources (see text for details). Curve $a$, in brown,
represents a stellar orbit at the 4:1 resonance, and curve $b$ an orbit at the 6:1 resonance. Curve $c$ is not identified 
in terms of resonances, since it is a linear fit to the observed points. The other lines (two in red, one in green) 
are logarithmic spirals fitted to the data.}
\label{fig10} 
\end{figure} 

\begin{figure*}
\centering
\includegraphics[width=18cm]{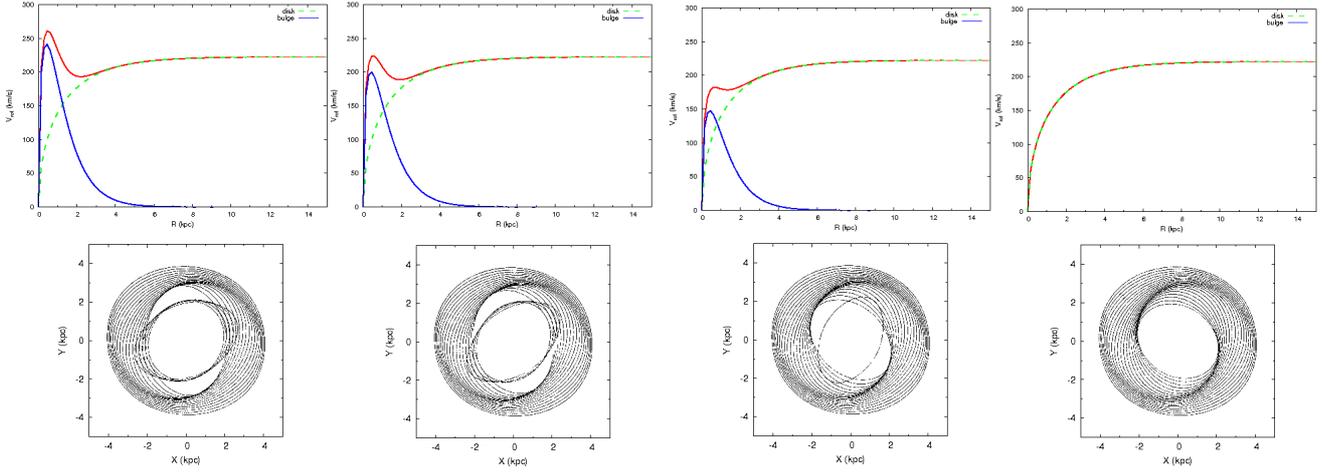}  
\caption{Rotation curves and corresponding inner orbits for different strengths of the bulge. At the top are the rotation curves for different 
values of $f_b$ (see text). From left to right, the parameter $f_b$ is 3.2, 2.2, 1.2, and 0, respectively. Below are the orbits associated 
with each rotation curve. The blue line is  the bulge contribution, the green line is the disk contribution, and the red line is the 
bulge $+$ disk contribution.}
\label{fig9} 
\end{figure*}

\section{Results}
\label{res}
\subsection{ A model for our Galaxy}
\label{mfg}

Throughout this work, we have adopted the parameters that appeared to us to correspond to our
Galaxy, like those presented in Table 1, and the rotation curve. In the next sections, 
we present both general results, which are valid in principle for other spiral galaxies,
and specific results, which  depend on the exact choice of parameters.
One of the most important choices when we apply a model to a galaxy is that of the
pattern speed, or equivalently, if the rotation curve is known, of the corotation radius. 

In the last few decades, many papers have pointed to a corotation radius of the Galaxy situated
close to the orbit of the Sun (\citealt*{Marochnik1972}, \citealt*{Creze1973}, \citealt*{Mishurov1999},
among others). An additional argument is that the classical theory of spiral arms tells us that the arms 
exist between the Inner and Outer Lindblad resonances and that the corotation falls roughly midway
between these resonances. The data from the literature collected by \cite{Scarano2012} supports the
theory, showing that the corotation occurs about midway between extremities of the 
region where the arms are seen. In our Galaxy, the spiral arms extend from about 3 kpc to 13 kpc
\citep[e.g.,][]{Russeil2003}, which would situate the corotation at about 8 kpc.  
Recently, a few direct methods allowed the decrease of the uncertainty on the corotation radius to a few 
hundred parsecs. One example is the determination of the pattern speed by integrating the orbits of
open clusters back to their birthplaces. As shown by \cite{Dias2005}, the result, $R_{CR} = (1.06 \pm 0.08) R_{sun}$,
does not have any strong dependence on the adopted rotation curve. A second direct observation
is the position of the ring-shaped HI void at corotation (\citealt{Amores2009}), and a third one 
the position of the square-shaped spiral arm associated with the 4:1 resonance \citep{Lepine2011a}.
The existence of a gap in the distribution of open clusters and of Cepheids, along with a
step in the metallicity distribution, is naturally explained in terms of the corotation resonance
\citep{Lepine2011b}. What we call a direct method is one that does not involve any complex model 
with uncertain parameters and therefore generates robust results. This is not the case of n-body 
simulations, or chemical evolution models. Based on the results above, we adopt $R_{CR}= 8.4 \pm 0.2$ 
kpc; accordingly, with Eq. \ref{galcurve}, we find $\Omega_{p} = 23$ km s$^{-1}$ kpc$^{-1}$.  

In the present work, we have not investigated the spiral structure beyond corotation, a task left for
future work. As we discuss in the next sections, there is a change in the nature of the arms at
the 4:1 resonance. Only between the ILR and that resonance are the arms explained by the crowding of stellar
orbits.

There are models of the spiral structure that are in conflict with the corotation radius
that we advocate. For instance, model of the inner Galaxy by \cite{Dehnen2000} suggests 
that the radius of the Outer Lindblad Resonance (OLR) of the Galactic bar lies in the vicinity of
the Sun. His model does not belong to the category of direct methods, as we defined above, since it is
a simulation and by nature, based on  many approximations and uncertain parameters.
The author does not claim that he presents a precise determination of the OLR of the bar
and suggests that this radius is $\approx$ 6-9 kpc.
A recent review of pattern speeds in the Galaxy is given by \cite{Gerhard2011}. He recognizes that the pattern
speed of the spiral structure is about 25 km s$^{-1}$ kpc$^{-1}$, but favors the idea that the bar rotates at a higher velocity. 
The review does not mention the work of \cite{Ibata1995}, who determined the rotation speed of the bulge, which was 
found to be the same that we adopt for the spiral arms. In a study of the morphology of bars in spiral galaxies, 
\cite{Elmegreen1985} concluded that in late-type spirals the bar may extend only to the ILR. This is the hypothesis that
we adopt to compare our model with the Galaxy, based on the fact that the solar neighborhood contains
spiral arms only.

\subsection{The structure  between the ILR and the 4:1 resonance}

Many studies of self-consistency have been performed \citep{Contopoulos1988,Patsis1991,Amaral1997,Pichardo2003,Martos2004,Antoja2011}. \cite{Contopoulos1986} already studied the response density and concluded that strong 
spirals do not exist beyond the 4:1 inner resonance. This result was confirmed by the other works and also 
by the present work. In Fig. \ref{fig8} we show the locii of maximum response density of our model. It can be
seen that there is a very good self-consistency of the shape of the spiral arms in the region inside the 4:1 resonance.
This self-consistency is particularly satisfactory in the case of our model, since a potential perturbation with a Gaussian 
profile generates a very similar response density and response potential. The fact that two well$-$behaved
logarithmic spiral arms extend about the 4:1 resonance is a robust result, which was encountered
by the various works mentioned above, although they use a variety of  axisymmetric potentials and perturbation
potentials. It also corresponds to an observed characteristic of spiral arms, since \cite{Elmegreen1995}
observed that most spiral galaxies  have two prominent symmetric arms in their inner regions, 
inside approximately 0.5 $R_{25}$, where $R_{25}$ was defined by them as the radius for which the surface brightness is 25 mag arcsec$^{-2}$. 
These authors suggest that the termination of these arms could coincide with corotation. This robust structure, with a similar aspect
in many spiral galaxies, is possibly the one that is able to determine the pattern speed. That the pattern speeds are found to have 
a distribution concentrated around 24 km s$^{-1}$ kpc$^{-1}$ \citep{Scarano2012}, in other words, they are not arbitrary.

We would like to call attention to an interesting  effect. One can see that the maximum response density 
starts to be out of phase with respect to the logarithmic spiral perturbation near the 4:1 resonance.
This suggests that the deviation from a perfect logarithmic spiral in galaxies can be understood naturally
as an effect of the 4:1 resonance. A consequence of this is the appearance of bifurcations. They happen
because when the response density starts to be out of phase with the imposed density, part of the matter
remains following the imposed density. This results in a weaker piece of arm. In galaxies with a structure similar
to that of the Milky Way, the appearance of bifurcations is possibly associated with the 4:1 resonance,
rather than with the corotation. This is not a new result, since it was already discovered from a theoretical approach,
for instance, by \cite{Patsis1994} and \cite{Patsis1997}. This is also a robust result because it does not depend
on the details of a given rotation curve. Observationally, the work of \cite{Elmegreen1995} 
also states that bifurcations start at the end of the two prominent symmetric arms.

 \subsection{Beyond the 4:1 resonance}
 
In our model, there is a clear change in the nature of the arms beyond the 4:1 resonance. 
There are no more arms produced by the concentration of stellar orbits, but only discrete
resonant orbits. How could a resonant orbit behave like a spiral arm? This may happen if stars
are captured within resonances, similarly to asteroids in the Solar system, for which the trapping 
mechanism was described by \cite{Goldreich1965}. For instance, in the asteroid belt, the 3:2, 4:3, and 1:1 
resonances with Jupiter are populated by clumps of asteroids. Similarly, the stars captured within resonances could produce
a local increase of density and turn it into a kind of arm.
If this model is correct, when applied to our Galaxy, it would predict that some of the arms seen
in the solar neighborhood have the form of resonant orbits.

Direct evidence of an arm with the shape of a 4:1  resonant stellar orbit was found 
by \cite{Lepine2011a} using as tracers molecular CS sources, Cepheids with a period longer than six days, and open clusters
with ages less than 30 Myr. The map obtained in that paper is reproduced in Fig. \ref{fig10}.
Here, the same symbol has been used for all the kinds of tracers, since we are only interested in the
general structure. It must be remembered that nowadays only in the Solar neighborhood we can have a detailed description of the spiral structure, since interstellar extinction does not allow 
open clusters and Cepheids to be observed at more than a few kpc.

Figure \ref{fig10} can be seen as a zoom of the upper part of Fig. \ref{fig8}, in the region around the Sun. One can 
observe the presence of a structure with an angle of about 90$\degr$ (labeled ``a'' and brown in the figure), which was identified by \cite{Lepine2011a} with the 4:1 resonance. The argument was not only the 
fit presented in this figure, but also much more the points observed along a straight portion of about 7 kpc in length,
in addition to the fact that the resonance is exactly where it is predicted to be, based on the known
pattern speed. Now, we also present a fit of an orbit at the 6:1 resonance to many of the observed points
(curve with label ``b'', in blue). This greatly resembles a predicted structure in Fig. \ref{fig8},
which passes close to the corner of the arm at the 4:1 resonance and close to the Sun, in addition to presenting an acute corner 
in the second quadrant of the Galaxy. 

Stable periodic orbits attract around them quasi-periodic orbits, as indicated by the contours of the islands
in the Poincar\'e's surface of section. It seems that in the region of the disk discussed here, quasi-periodic orbits
 are trapped around the stable periodic orbits of the 4:1 and 6:1 families, with very small deviations from the 
periodic orbits themselves. As a result the quasi-periodic orbits could be considered as building a 
``thick periodic orbit'' structure. These arms, which are composed of similar quasi-periodic orbits, would have very small
dispersions of velocities.

Finally, one can see in Fig. \ref{fig8}, just below the corner of the
4:1 resonance, an orbit that crosses it in two points. This resembles the arm-like 
structure with label C in Fig. \ref{fig10}. We consider that this observed structure is probably real, due to the large 
number of tracers that fall on it. This suggests an interesting concept. Since resonant orbits
can cross each other, it is also possible to have arm-like structures which correspond to them
crossing each other. This contrasts with the classical view of spiral arms, in which no such intersections
exist.

Of course, the interpretation we give of some local structures may not seem convincing. One possible
argument against it is that the resonant orbits do not seem to be uniformly populated by stars all around
them. However, the mechanism of the capture of stars by resonances deserves deeper study to determine if 
some regions of the orbits should present a larger population. 
The solar neighborhood  is rich in structures that have not been explained by any model. One could argue that they
are not explainable because they are due to stochastic events. The idea that we propose here, that
some of these structures correspond to resonant orbits, can be verified, since
at each point of such an orbit we have a particular velocity of the stars that are trapped into it.
However, this is not the scope of the present paper.  We did not make any adjustment 
of the  parameters of the model in order to fit observed structures. Parameters like $V_0$ (which affects the
 rotation curve), the exact depth and  location of the local minimum in the rotation curve, $\Omega_p$, and the width, depth, and pitch angle of the imposed perturbation could all have been submitted to minor variations. 
Our experience is that this is a lengthy task, with minor variations of the structures produced
and some redundancy (different combinations can give similar results). The comparison with the 
observed data is difficult in part due to its poor quality. Besides, some fitted
models  may look better in one region of the galactic plane but worse in another.  So, we also leave 
this task for future work.

Like the  previous works on self-consistent spiral structure, we did not extend the calculation of
stable orbits beyond the corotation radius. We expect to see resonant orbits as well, but the comparisons
with real arms will be even more difficult because the distances and shapes of the external arms of our Galaxy 
are poorly known. What is known, from an observational point of view, is that the spiral arms show strong
departures from logarithmic spirals in the external regions of  galaxies. In most cases, the arms look
more like a sequence of straight segments that roughly follow a spiral path \citep[e.g.,][]{Chernin1999}.

\subsection{The bulge and the bar}
\label{bb}

About $50\%$ of spiral galaxies have a bar, either strong or weak, and there are both kinematic and photometric
evidences that the Milky Way has a bar \citep[Sect.~6.5]{Binney2008}. \cite{Shen2010} argue that the Galactic
bulge is a part of the disk and suggest that the bulge (in reality, a boxy pseudobulge) of the Milky Way is
an edge-on, buckled bar that evolved from a cold, massive disk.

\begin{figure}
\hspace{-1cm}
\centering
\includegraphics[scale=0.24]{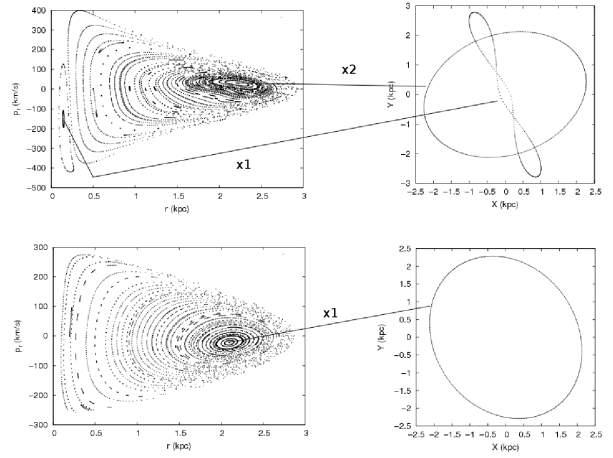}\centering    
\caption{Poincar\'e diagrams and their respective families $(x_1,x_2)$ of perturbed periodic orbits \citep[see][for details]{Contopoulos1975}. 
In both cases we set the energy at $R_c= 2.2$ $kpc$, and the perturbation parameters are the same as Table \ref{tab1}. The only difference  
is the bulge fraction $f_b$; at the top we have $f_b = 3.2$ and below we have $f_b=0$.}
\label{fig11} 
\end{figure}

In this section, we analyze the connection between the spherical central bulge and the inner stellar orbits. Figure 
\ref{fig9} shows the rotation curves from Eq. \ref{flatcurve}. The scale lengths for the disk and the bulge 
are fixed at $\epsilon_{d}^{-1} = 2$ kpc$^{-1}$ and $\epsilon_{b}^{-1} = 0.4$ kpc$^{-1}$, $V_{max} = 220$ km s$^{-1}$.
The parameter $f_b$, defined in Eq. \ref{flatcurve} as ``the bulge fraction'', is equivalent to a measure of
the ``bulge strength'', since the disk component is kept constant (only the bulge is varied), or even to a measure
of the bulge density, since its scale size is kept constant. We will refer to it as the bulge strength.
We varied  $f_b$ from 3.2 down to 0. Below each rotation curve in  Fig. \ref{fig9}, we
show the family of perturbed periodic orbits trapped around unperturbed periodic orbits from $R_c = 2$ up to $4$ kpc;
see Sect. \ref{is} for more details. For a given energy $h$, which correspond to a radius $R_c$, we may have more then one periodic orbit. 
The family of periodic orbits lies at the center of the island in the Poincar\'e diagram. We call it family $x_1$ if the center of the island
lies at R $\leq$ $R_c$, and family $x_2$ if R $\geq$ $R_c$, \citep[see][]{Contopoulos1975}.

The parameters for the spiral perturbation are the same as given in Table \ref{tab1}. The orbits in the inner region, almost perpendicular 
to the more external orbits, belong in Fig.\ref{fig9}, to the family $x_2$ of periodic orbit. They are more prominent where we have a strong central
bulge $(f_b = 3.2)$; as the bulge weakens, the families of orbits $x_2$ disappear, as we can see in Fig. \ref{fig11}.
In the case where $f_b \neq 0$, we have two families of stable orbits $(x_1, x_2)$. In Fig. \ref{fig9}, we plot the 
orbits corresponding to families for which we have only  small deviations from $R_c$ and those with a realistic
distribution of velocities. As we can see in Fig. \ref{fig11} on top, the family $x_1$ has a large 
deviation from $R_c$ with a velocity in the order of $p_r \approx -150$ km s$^{-1}$ generating a very eccentric orbit,
while the family $x_2$ is close to $R_c$ with $p_r \approx 20$ km s$^{-1}$; thus, we expect that it would be the
favorite family to populate this region. It is clear that strong bulges  favor the $x_2$ family in the inner regions
of the Galaxy, while without a bulge we have only the family $x_1$.

A visual inspection of the orbits associated with different rotation curves shows elongated stellar orbits in the inner
regions of the disk for rotation curves with the presence of a peak near the center, suggesting the formation of a
bar with a half-size  $\sim3$ kpc. This model shows the existence of a correlation between the formation of a bar
and the magnitude of the central bulge. The peak in the rotation curve becomes visible  when $f_b$ is relatively
large (3.2 or 2.2). A strong bulge generates a prominent velocity peak in the rotation curve and modifies the inner stellar orbits. 
In other words, the strength of the bulge has an influence on the existence of a bar. This bar appears as a consequence of the same
potential perturbation that we introduced for the spiral arms, so that the bar and the spiral arms have the same
rotation velocity. The type of bar that we are discussing here corresponds to what is understood as a weak bar. In principle, this would seem to be more compatible with the model of spiral arms presented in this work, in
which no effect of a bar is taken into account in the large ``spiral'' region that we described.
A bar is weak if it generates only a small fraction of the total gravitational field around it. In such systems,
the stellar orbits can be described by the epicyclic theory plus a weak driving force due to the periodic motion of
the bar. In our Galaxy, the mass associated with the spherical bulge is much larger than that associated 
to the bar, as deduced from infrared luminosity (e.g., \cite{Lepine2000}). One can even expect that at a 
distance of the order of 1 kpc from the center, outside the bulge but under its strong influence, the potential 
is Keplerian, with a dominant mass at the center. In such a potential, the orbits are elliptic, with the center
of the Galaxy at one focus of the ellipse, like the orbits of comets in the solar system. In epicycle theory,
these are 1:1 orbits. Two families of such orbits aligned along a straight line, one on each side of the bulge,
would be in part responsible for the bar. The alignment of such elongated, one-sided orbits with the inner extremity
of the spiral arms would occur naturally, since stars travel slowly when they are at the apogalactic parts of their orbits
and this facilitates synchronization. Of course, we are not presenting here a complete model of a weak bar,
but only arguing that the existence of a weak bar in our Galaxy should not be rejected. If we consider that
the origin of the spiral structure is related to the interaction with an external galaxy, it is more logical to 
think that the bar is induced by the spiral structure, and not vice versa.
An interesting possibility is that since the only imposed perturbation here is that of the spiral arms, our results
support the idea that a pseudobulge could have evolved from a perturbed disk \citep{Shen2010}.

We must add to the list of similarities between our model and the real Galaxy the agreement, in Fig. \ref{fig8}, 
between the orientation of the elongated inner orbits, where the spiral arms seem to begin, with respect to the Sun.
This is very similar with the observed angle between the Sun and the extremities of the bar \citep[see, e.g.,][]{Lepine2011a}. 

\section{Conclusions}
\label{con}

We have presented a new description of the spiral structure of galaxies, based on the interpretation of the arms 
as regions where the stellar orbits of successive radii come close together, producing large stellar 
densities. In other words, the arms are seen as grooves in the potential energy distribution. Such an approach 
is not new \citep[e.g.,][]{Contopoulos1986, Amaral1997}. The innovation is that the potential energy 
perturbation that we have adopted is itself like a groove that follows a logarithmic spiral, with a Gaussian 
profile in the direction perpendicular to the arms. This represents an important step in the direction of 
self-consistency, since this potential perturbation generates, by means of the stellar orbits, spiral arms
with a similar profile. In previous classical models, the potential perturbation was represented by a 
sine function (or a sum of two sine functions, if four arms are presented) in the azimuthal direction,
but the response potential resembled the groove-like one of the present work, so that the self-consistency was poorer.
One of the classical models \citep{Pichardo2003} does not use a sine function for the imposed potential, 
but its complexity does not allow an easy check of self-consistency.

A new parameter appears in our description, allowing us to control the width of the arms. We found a relation 
between the density contrast between the arm  and inter-arm region, and the perturbation amplitude, under the 
assumption of an exponential disk. 

We confirm the conclusion of previous works that the 4:1 inner resonance is a fundamental structure of the disk
and that similar strong arms do not appear beyond this resonance. This result is supported by observations, 
since it is known that most spiral galaxies have two symmetric prominent arms in their inner regions 
\citep{Elmegreen1995}.

The model is applied to our Galaxy, using a description of the axisymmetric potential similar to
that derived from the observed rotation curve, along with the known pattern rotation speed and pitch angle.
The range of values for the perturbation amplitude compatible with observational evidences of the contrast density
is $400 - 800$ km$^2$ s$^{-2}$ kpc$^{-1}$. Using a density contrast of about $20\%$, an 
accepted average value, the perturbation amplitude $\zeta_0$, is equal to 600 km$^2$ s$^{-2}$ kpc$^{-1}$.
It is found that the orbits in the region between the 4:1 resonance and corotation do not reinforce spiral arms like
in the inner regions, but only features similar in shape with the periodic orbits existing in this region.
If the model can validly explain the spiral arms in the solar neighborhood, then the
dynamics in that region is determined by the motion of stars in quasi-periodic orbits.

Interestingly, a number of observed structures are very similar to the predicted resonant orbits.
The similarities of the model with the Galaxy include the orientation of the bar, the size and orientation
of the 4:1 resonance orbits as revealed by \cite{Lepine2011a}, and possibly a resonant orbit with outward 
peaks or ``corners'' and inverted curvature between the corners, situated close to the Sun.

The idea that the spiral structure can be self-consistent (a certain perturbing potential gives rise to stellar 
orbits that reproduce this perturbation) points towards a long-lived structure. This view is in conflict with 
some recent interpretation of the spiral arms as transient features \citep[e.g.,][]{Sellwood2011}, but consistent
with the argument that the spiral structure must be long lived based on the observation of the metallicity step 
at corotation \citep{Lepine2011b}.

Another interesting result of the present model is that a bar naturally appears, without imposing any special 
condition except that the spiral perturbation potential extends a little inwards of the ILR. We can see elongated orbits 
in the central regions of the Galaxy, which we identify as being a weak  bar. We found a correlation between the 
presence of the elongated central orbits and the mass of the bulge. The bulge of our Galaxy is a relatively 
massive one, as revealed by the peak in the rotation curve near the center, and consequently can generate a weak bar, according to this model. This bar would have the same pattern speed of the spiral arms. 

We are conscious that there are many works in the literature, that are in conflict with our results. They propose, for instance, 
different corotation radii and different values for the pattern speed, the existence of several pattern speeds in our Galaxy, or that the arms are transient structures. Another idea is 
that the bar rotates with a different speed relative to the spiral arms. There is also a fashionable artistic view of 
the spiral structure offered by a NASA site, often reproduced in scientific works, which presents the Galaxy as 
a perfect logarithmic spiral up to large radii. Such a structure could not be a transient one. Still others
consider that swing amplification underlies the basic physics of the spiral arms. To the many contradictory ways of understanding  the Galactic structure, our group of research adds
one more possibility. We see the arms as relatively long-lived grooves in the gravitational potential of the disk, whose 
shape is primarily determined by the stellar orbits and not by shock waves in the interstellar medium.
The spiral structure is self-consistent in terms of arm shapes in the range of radius between the inner Lindblad resonance
and the 4:1 resonance, so that this region is probably the one that guarantees the stability of the whole structure
and imposes its pattern speed. Outside this region, many observed structures have shapes coinciding with
predicted orbits at resonances. Since our model predicts the existence of resonant orbits with reverse
curvature between corners pointing outward (as in our Fig. \ref{fig8}), it is not surprising to observe
structures with similar characteristics in our Galaxy. Similarly, it is not too surprising to find
arm-like structures crossing each other. The expected response of the gas of the disk is to get trapped into
the grooves and to flow along them.

\begin{acknowledgements}
The work was supported by the Conselho Nacional de Desenvolvimento Cient\'ifico e Tecnol\'ogico.  We thank
Dr. P. Grosb{\o}l for many helpful comments and to the anonymous referee for useful comments. 
We also thank Vinicius C. Busti for many useful comments.   

\end{acknowledgements}

\begin{appendix}
\section{Deduction of the Gaussian potential following a logarithmic spiral}
\label{dop}

A Gaussian in two dimensions with equal standard deviation $\sigma = \sigma_x = \sigma_y$ is given by

\begin{equation}
\label{gxy}
{
f(x,y) = Ae^{\frac{-(x-x_{0})^2 - (y-y_0)^2}{2\sigma^2}}.
} 
\end{equation}

In polar coordinates we have

\begin{equation}
\label{xx}
{
(x-x_{0})^2 = R^2\cos^2(\theta) - 2R\, R_0\cos(\theta)\cos(\alpha) + R_{0}^2\cos^2(\theta),
} 
\end{equation} 

\noindent and

\begin{equation}
\label{yy}
{
(y-y_{0})^2 = R^2\sin^2(\theta) - 2R\,R_0\sin(\theta)\sin(\alpha) + R_{0}^2\sin^2(\theta),
} 
\end{equation} 

\noindent where $R_0$ = $\sqrt{x_0^2 + y_0^2}$ gives the position of the logarithmic spiral. If we fix the 
radii and just look at the variation in the azimuthal direction, Eqs. \ref{xx} and \ref{yy}, become

\begin{equation}
\label{xx2}
{
(x-x_{0})^2 = R^2\cos^2(\theta) - 2R^2\cos(\theta)\cos(\alpha) + R^2\cos^2(\theta),
} 
\end{equation} 

\begin{equation}
\label{yy2}
{
(y-y_{0})^2 = R^2\sin^2(\theta) - 2R^2\sin(\theta)\sin(\alpha) + R^2\sin^2(\theta).
} 
\end{equation}

Adding the two equations and using the law of cosines we have

\begin{equation}
\label{xy2}
{
(x-x_{0})^2 + (y-y_{0})^2 = 2R^2( 1 - \cos(\theta - \alpha)).
} 
\end{equation}

\noindent Taking Eq. \ref{xy2} and putting into Eq. \ref{gxy}, this becomes

\begin{equation}
\label{gp}
{
f(R,\theta) = Ae^{\frac{-R^2}{\sigma^2}\left[1 - \cos(\theta - \alpha)\right]}.
} 
\end{equation}

Now we can set the angle $\alpha$ as a logarithmic function, which defines the spiral arm:

\begin{equation}
\label{alf}
{
\alpha = \frac{ln(R/R_i)}{\tan(i)} + \gamma,
} 
\end{equation}

\noindent but Eq. \ref{gp} describes only one arm. However, it is simple to generalize for $m-arms$. We simply multiply the term
$(\theta - \alpha)$ inside the cosine, in Eq. \ref{gp}, by $m$:

\begin{equation}
\label{mgp}
{
f_m(R,\theta) = Ae^{\frac{-R^2}{\sigma^2}\left\{1 - \cos[m(\theta - \alpha)]\right\}}.
} 
\end{equation} 

This is the final equation to describe $m-arms$.

\end{appendix}

\begin{appendix}
\section{Thin-disk approach for the new potential}
\label{apB}

The Poisson's equation for a thin disk is

\begin{equation}
\label{poisson}
{
\nabla^{2}\Phi_1 = 4\pi G \Sigma \delta(z),
} 
\end{equation}

\noindent where we can write the potential as follows:

\begin{equation}
{
\Phi_{1}(R,\theta,z) = \Phi_{10}(R,\theta) e^{-|kz|}.
} 
\end{equation}
   
Therefore the Poisson's equation has the form

\begin{equation}
\label{poisson2}
{
\Sigma \delta(z)= \frac{1}{4 \pi G} (e^{-|kz|}\nabla_{R,\theta}^{2}\Phi_{10} + \Phi_{10}\frac{\partial^{2}e^{-|kz|}}{\partial z^2}),
} 
\end{equation} 

\noindent where $\nabla^{2}_{R,\theta}$ is the Laplacian in polar coordinates. Integrating both sides of this equation from $z = -\infty$ to $z=\infty$,
the second term on the right-hand side of this equation disappears, then Eq. \ref{poisson2} becomes

\begin{equation}
\label{sigpois}
{
\Sigma = \frac{1}{2 \pi G} (\frac{1}{k}\nabla_{R,\theta}^{2}\Phi_{10}).
}
\end{equation}

For the potential $\Phi_{10}$ equal to
 
\begin{equation}
 {
\Phi_{10} = - \zeta_0\, Re^{\frac{-R^2}{\sigma^2}[1 - \cos\chi] - \varepsilon_s R},
}
\end{equation}

\noindent with $\chi = m(\theta - \alpha)$, and $\alpha$ given by Eq. \ref{alf}. 

Solving $\nabla_{R,\theta}^{2}\Phi_{10}$ we have that

\begin{equation}
\label{poisson3}
 {
\nabla_{R,\theta}^{2}\Phi_{10} = \zeta_0 e^{-\psi}(\phi_1 - \phi_2),
}
\end{equation}

\noindent where

\begin{equation}
 {
\psi = \frac{R^2}{\sigma^2}\left(1 - \cos\chi\right) + \varepsilon_s R,
}
\end{equation}

\begin{equation}
{
\phi_1 = 3\frac{\partial\psi}{\partial R} + R\left[\frac{\partial^2\psi}{\partial R^2} - 
\left( \frac{\partial\psi}{\partial R} \right)^2 \right],
}
\end{equation}

\noindent and

\begin{equation}
{
\phi_2 = \frac{1}{R} \left[ 1 + \left(\frac{\partial\psi}{\partial \theta} \right)^2
- \frac{\partial^2\psi}{\partial \theta^2} \right].
}
\end{equation}

The derivatives of $\psi$ are

\begin{equation}
{
\frac{\partial\psi}{\partial R} = \frac{2R}{\sigma^2}\left( 1 - \cos\chi - \frac{m}{2\tan i}\sin\chi\right) + \varepsilon_s,
}
\end{equation}
 
\begin{equation}
{
\frac{\partial^2\psi}{\partial R^2} = \frac{1}{\sigma^2}\left( 2 + \left(\frac{m^2}{\tan^2 i} - 2 \right)\cos\chi - 
\frac{3m}{\tan i}\sin\chi\right),
}
\end{equation}

\begin{equation}
{
\frac{\partial\psi}{\partial \theta} = \frac{mR^2}{\sigma^2}\sin\chi,
}
\end{equation}

\noindent and

\begin{equation}
{
\frac{\partial^2\psi}{\partial \theta^2} = \frac{m^2R^2}{\sigma^2}\cos\chi.
}
\end{equation}

For a fixed $R$ the maximum density occurs when $\cos\chi = 1$, $\sin\chi = 0$. Assuming a small pitch angle $i$,
the most important terms contains $\tan^2 i$ in the denominator. Thus Eq. \ref{poisson3} is reduced to

\begin{equation}
 {
\nabla_{R,\theta}^{2}\Phi_{10} = \frac{\zeta_0 m^2 R}{\sigma^2\tan^2 i}e^{-\varepsilon_s R}.
}
\end{equation}

Thus the maximum density computed from Eq. \ref{sigpois} is

\begin{equation}
 {
\Sigma_{max} = \frac{\zeta_0}{2\pi G k}\frac{ m^2 R}{\sigma^2\tan^2 i}e^{-\varepsilon_s R},
}
\end{equation}

\noindent from the classical spiral density wave theory \citep{Lin1964}. The spiral arms were modeled with a cosine function where the wavenumber
$k$ appears, and it is equal to  $k = \frac{m}{R |\tan i|}$ for a logarithmic description. We know that our expression is not a pure cosine 
law  anymore. However, we can locally think of the Gaussian density as a wave for which the amplitude tends to zero far away, so that we can keep the 
classical definition of $k$ and use it on the equation above. Taking $m=2$ we have 

\begin{equation}
 {
\Sigma_{max} = \frac{\zeta_0}{\pi G}\frac{R^2}{\sigma^2|\tan i|}e^{-\varepsilon_s R}.
}
\end{equation}

If we chose $\sigma = R$, we recover the classical formula from \cite{Lin1969} 

\begin{equation}
{
\Sigma_{max} = \frac{\zeta_0}{\pi G}\frac{1}{|\tan i|}e^{-\varepsilon_s R}.
}
\end{equation}

Now the density contrast not only depends on the perturbation amplitude $\zeta_0$, but also has a dependence with $\sigma$, which 
describes the width of the arms. In our model, we can set $\sigma$ as a constant or a function of R.

\end{appendix}

\bibliographystyle{aa}
\end{document}